\documentclass[12pt,letterpaper,notitlepage]{article}

\usepackage{epsfig}
\usepackage{amsmath}
\usepackage{cite}

\newcommand{\tev}{\,\, \mathrm{TeV}}
\newcommand{\gev}{\,\, \mathrm{GeV}}

\newcommand{\gesim}{\,{_{\textstyle>}\atop^{\textstyle\sim}}\,}

\begin{document}

\title{High Energy $WW$ Scattering at the LHC with~the~Matrix~Element~Method}

\author{A.~Freitas$^1$ and J.~S.~Gainer$^{2,3,4}$}

\date{}

\maketitle

{\centering
$^1$ Pittsburgh Particle-physics Astro-physics \& Cosmology
    Center (PITT-PACC),\newline  Department of Physics \& Astronomy,
    University of Pittsburgh, Pittsburgh, PA 15260, USA \\[1ex]
$^2$ High Energy Physics Division, Argonne National Laboratory,
Argonne, IL 60439, USA \\[1ex]
$^3$ Department of Physics \& Astronomy, Northwestern University,
Evanston, IL 60208, USA \\[1ex]
$^4$ Physics Department, University of Florida, Gainesville, FL 32611,
USA\footnote{Current address.}\\
}

\vspace{2em}

{\centering\small
NUHEP--TH/12--12,~
ANL--HEP--PR--12--104\\
}

\vspace{2em}

\begin{abstract}
Perhaps the most important question in particle physics today is
whether the boson with mass near 125 GeV discovered at the Large Hadron Collider
(LHC) is the Higgs Boson of the Standard Model. Since a
particularly important property of the Standard Model Higgs is its
role in unitarizing $W_{\rm L}W_{\rm L}$ scattering, we study the ability of the
LHC to probe this process in the case of same-sign W pair production.
We find that the use of the Matrix Element Method increases the
significance with which the Higgs sector can be probed in this
channel. In particular, it allows
one to distinguish between a light and heavy SM Higgs in this
channel alone with a high degree of significance, as well as to set
important limits in the parameter space of the Two Higgs Doublet
Model and the Strongly-Interacting Light Higgs Model with less than
200~fb$^{-1}$ at the 14~TeV LHC, thus providing crucial
information about the putative Higgs boson's role in electroweak
symmetry breaking.
\end{abstract}


\clearpage


The mechanism of electroweak symmetry breaking (EWSB) can be tested most
directly in high energy vector boson scattering. In fact, the tree-level
amplitudes for scattering of longitudinally polarized $W_{\rm L}W_{\rm L}$,
$Z_{\rm L}Z_{\rm L}$ and $W_{\rm L}Z_{\rm L}$ pairs involving only vector bosons
grow unboundedly with energy until they violate the unitarity limit. This
unphysical growth must be canceled by the dynamics of EWSB. For instance, in the
Standard Model (SM), this is achieved by diagrams involving Higgs boson
exchange. After the recent discovery of a Higgs-like resonance with a mass
of 125--126~GeV \cite{higgs1} it will be essential to explicitly test whether
this particle, other new physics, or a combination of the two are
responsible for the unitarization of vector boson scattering.

In $pp$ collisions, this question can be studied through processes of the type
$pp \to jjV_1^*V_2^* \to jjV_1V_2$, where $V_i^{(*)}$ stands for a
(off-shell) $W$- or $Z$-boson. Since these are four-body processes, the cross
sections are small: even for $\sqrt{s}=14\tev$, the typical signal rates are
${\cal O}$(fb) or less. Furthermore, only a fraction of this rate is contributed
by  longitudinal vector bosons, so the analysis of such processes at the LHC is
very challenging. Consequently, much effort has been invested in studying
high energy vector boson scattering and the
improvement of signal selection cuts
\cite{history,phantom11,wwpol}. Most of these papers focused on counting signal
events or analyzing individual distributions (such as the $VV$ invariant mass
distribution). However, additional information may be gleaned from various
angular correlations, which are sensitive to the details of the
unitarity-restoring dynamics.

The Matrix Element Method (MEM) \cite{matrix,memtev1} is a
promising approach for comprehensively taking into account all information
from an event and thus maximizing the sensitivity. It provides an algorithm for computing the
likelihood that a given experimental event sample agrees with a theoretical
model, using parton-level matrix elements and a set of input
parameters. For each single event, with observed momenta $\textbf{p}_i^{\rm
vis}$, the likelihood that it agrees with a given model and set of model
parameters $\alpha$ is defined as
\begin{equation}
{\cal P}(\textbf{p}^{\rm vis}_i|\alpha) = \frac{1}{\sigma_\alpha} 
\sum_{k,l} \int dx_1 dx_2 \, \frac{f_k(x_1)f_l(x_2)}{2sx_1x_2}
 \;\biggl [ \prod_{j \in \text{inv.}} \int
 \frac{d^3 p_j}{(2\pi)^3 2E_j} \biggr] |{\cal M}_{kl}(p_i^{\rm vis},p_j;\alpha)|^2.
 \label{eq:mem1}
\end{equation}
Here $f_k$ and $f_l$ are the parton distribution functions for the initial-state
partons $k$ and $l$, respectively, ${\cal M}_{kl}$ is the theoretical matrix element,
and $\sigma_\alpha$ is the total cross section, computed with the same matrix
element. The momenta $p_j$ of any invisible particles, such as neutrinos, are
integrated over the available phase space. The combined logarithmic likelihood
for a sample of $N$ events approximately converges to a $\chi^2$ distribution:
\begin{equation}
\chi^2 = 
-2\ln({\cal L}) = 
-2\sum_{n=1}^N \ln {\cal P}(\textbf{p}^{\rm vis}_{n,i}|\alpha)
,
\label{eq:logl}
\end{equation}
where $\textbf{p}^{\rm vis}_{n,i}$ are the measured momenta of the $n$th
event.

By searching for the maximum of the likelihood, one can discriminate between
several models and/or determine the parameters of a given model. The method is
particularly powerful for the measurement of processes with low event yield
and/or unreconstructible event kinematics due to the presence of invisible
particles in the final state. Typical examples are top quark physics at the
Tevatron \cite{memtev1,memtev2,singletop}, Higgs searches
\cite{memhiggs,memnlo}, and the production of dark matter
particles at the LHC \cite{lhcll,madweight}. 

This letter reports on the application of the MEM to the process $pp \to
jjW^+W^+ \to jj\ell^+\ell^{\prime+}\nu_\ell\nu_{\ell'}$, where
$\ell^{(\prime)}=e,\mu$ and $j$ denotes a light quark jet. While this channel
has a rather small cross section, due to the restriction to only positive charge
and leptonic decay channels of the $W$ bosons, it has the advantage of being
experimentally clean and having low background. As a result, its sensitivity can
be competitive with other vector boson scattering channels \cite{phantom11,wwpol}.
An analysis of the $W^+W^-$ channel, which has a larger signal rate, with the
MEM is left for a future publication.

For the likelihood calculation and the cross-section normalization in
eq.~\eqref{eq:mem1}, the complete set of tree-level diagrams for the partonic 
processes $q\bar{q} \to q'\bar{q}'W^+W^+ \to
q'\bar{q}'\ell^+\ell^{\prime+}\nu_\ell\nu_{\ell'}$ ($q,q' = u,d,s,c$) have been
included, $i.\,e.$ all diagrams with on-shell intermediate $W$ bosons. Besides
the contribution from longitudinal $W^+W^+$ scattering, this set also includes
the irreducible background from all other diagrams leading to the same final
state, $q'\bar{q}'W^+W^+ $. It has been shown that the difference
between considering only these diagrams and considering the full
process $q\bar{q} \to  q'\bar{q}'\ell^+\ell^{\prime+}\nu_\ell\nu_{\ell'}$,
including off-shell and t-channel $W$ exchange, is negligible for high energy
$WW$ scattering \cite{wwpol}.  The analysis is performed at the
parton-level---but note that the inclusion of transfer functions for jet
smearing and initial-state radiation is
conceptually straightforward \cite{memtev1,memtev2,memisr} (see also
\cite{memnlo}), although it 
substantially increases the computing time.

Signal events are defined through a set of preselection cuts:
\begin{equation}
\begin{aligned}
&p_{\rm T,\ell} > 20\gev, & &|\eta_\ell| < 2.5, \\
&p_{\rm T,j} > 30\gev, & &|\eta_j| < 5, 
& &\Delta R_{jj,\ell j,\ell\ell} > 0.4 \\
&|\eta_{j_1}-\eta_{j_2}| > 4, & &|\eta_j| > 1, 
& &\!m_{j_1j_2} > 100\gev, \\
&m_{\ell j} > 190\gev.
\end{aligned}
\label{eq:cuts}
\end{equation}
Here $p_{{\rm T},i}$ and $\eta_i$ are the transverse momentum and
pseudorapidity, respectively, of a final-state object $i=\ell,j$; $m_{ij}$ is
the invariant mass of the two objects $i$ and $j$; and $\Delta R_{ij} \equiv
\sqrt{(\eta_i-\eta_j)^2+(\phi_i-\phi_j)^2}$ quantifies the separation of two
objects in the plane of pseudorapidity and azimuthal angle. The first two lines
in eq.~\eqref{eq:cuts} describe the general detector acceptance and object
isolation cuts. The third line identifies the typical vector boson fusion
topology, where the two jets are expected to go in the forward and backward
directions, respectively\footnote{These cuts are kept relatively loose since
their primary purpose is to reduce the number of input event to a reasonable
amount. Further improvement of the signal significance is left to the MEM.}.
The last line removes background from $t\bar{t}$
production, which can produce an apparent same-sign lepton signal due to the
small, but non-negligible, probability for lepton sign misidentification in the
detector~\cite{wwpol}. The invariant mass of a lepton-jet pair originating
from top quark decay is bounded from above by the top mass $m_{\rm t}$, so the
requirement that the invariant mass of any lepton and jet is sufficiently larger
than $m_{\rm t}$ eliminates that background, while about 25\% of the signal
events are retained.

The determination of the MEM weights ($i.\,e.$ the numerator in
eq.~\eqref{eq:mem1}) and cross section normalization factors (denominator in
eq.~\eqref{eq:mem1}) has been carried out with a specialized private
code\footnote{The code is available upon request from the authors.}, using 
diagrams generated with {\sc FeynArts 3.3} \cite{feynarts}. As a cross-check,
the results have been cross-checked against {\sc MadGraph/MadEvent 5}
\cite{madgraph} and {\sc MadWeight 2.5} \cite{madweight} and good
agreement has been found\footnote{It was difficult to reach the
required precision for the cross section values with {\sc MadGraph/MadEvent}
within a reasonable amount of computing time, so the results presented here are
based on our private code.}. {\sc
MadEvent} has also been used for the generation of the simulated
``experimental'' events that are fed into the MEM fit. Throughout
this paper, the ``experimental'' events sample is based on the SM with $m_{H,\rm
ref} = 125\gev$. The cuts \eqref{eq:cuts} have been consistently applied both to
the event generation and weight normalization.

In the following the analysis of the process $pp \to jjW^+W^+ \to
jj\ell^+\ell^{\prime+}\nu_\ell\nu_{\ell'}$ with the MEM is presented for
three characteristic models: the Standard Model (SM), the Two Higgs Doublet
Model (THDM), and the Strongly Interacting Light Higgs (SILH) Model.

\begin{figure}[t]
\centering
\epsfig{figure=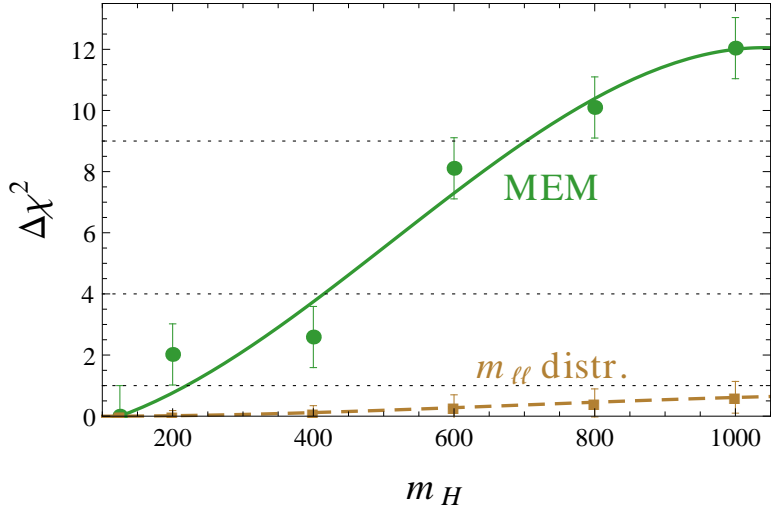, width=4in}%
\vspace{-1ex}
\caption{Statistical discrimination between different SM Higgs masses from
analyzing the process $pp \to
jjW^+W^+ \to jj\ell^+\ell^{\prime+}\nu_\ell\nu_{\ell'}$ using the MEM (circles)
vs.\ the di-lepton mass distribution (squares). The results are based
on 100 events simulated for $m_{H,\rm ref}=125\gev$ and $\sqrt{s}=14\tev$. The
error bars indicate the uncertainty from the event statistics.}
\label{fig:sm}
\end{figure}

\paragraph*{SM:} For $m_H = 125\gev$,
the signal cross section for $\sqrt{s}=14\tev$ after the preselection cuts is
0.590~fb, resulting in a signal yield of 100 events with an integrated
luminosity of 170~fb$^{-1}$. Assuming this number of signal events, the results
of the MEM likelihood fit are shown in Fig.~\ref{fig:sm}. Specifically, if one
wants to test whether the unitarization of $WW$ scattering is facilitated by a
light Higgs boson with $m_H = 125\gev$ or a heavy Higgs boson with $m_H =
1000\gev$, these two hypotheses could be distinguished with a statistical
significance of more than three standard deviations. 

For comparison, Fig.~\ref{fig:sm} shows the
statistical discrimination obtained from analyzing the $m_{\ell\ell}$
distribution as suggested $e.\,g.$ in Ref.~\cite{phantom11}. When using two bins
in the range $m_{\ell\ell} \in [0,1000]\gev$ and a sample of 100 events, the
significance stays below one standard deviation for Higgs masses up to 1~TeV. 
With larger numbers of bins the discriminative power is even lower.

Of course, in light of electroweak precision tests and the evidence for a 125-GeV
scalar at LHC \cite{higgs1}, 
the SM with a heavy Higgs boson is effectively excluded. Nevertheless it is illustrative
to discuss this scenario as an simple example that tangibly highlights the
difference between the MEM and a traditional analysis strategy.

\begin{figure}[t]
\centering
\epsfig{figure=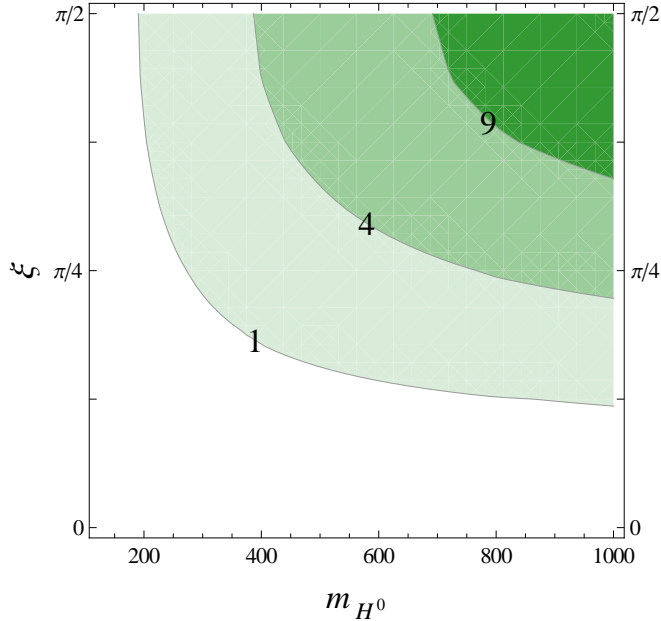, width=3.5in}
\vspace{-1ex}
\caption{Projected constraints on the THDM from the MEM analysis of the process
$pp \to jjW^+W^+ \to jj\ell^+\ell^{\prime+}\nu_\ell\nu_{\ell'}$, as function of
the heavy Higgs mass $m_{H^0}$ and the mixing angle $\xi$ (the light Higgs mass
is fixed to $m_{h^0}=125\gev$). The contour lines indicate $\Delta
\chi^2=1,4,9$. The results are based on 100 events simulated for
$\xi=0$ and $\sqrt{s}=14\tev$.}
\label{fig:2h}
\end{figure}

\paragraph*{THDM:}
A more realistic scenario is given by the THDM with a
light CP-even Higgs boson ($h^0$) of mass $m_h = 125\gev$, which explains the
observed Higgs signal, and a heavy CP-even Higgs ($H^0$) with
unconstrained mass $m_H$ \cite{thdm}. Assuming CP conservation, the two physical states
$h^0$ and $H^0$ are mixtures of the CP-even components of the two Higgs
doublets, $H^0_1$ and $H^0_2$:
\begin{equation}
\begin{aligned}
h^0 &= \cos\alpha\; H^0_1 - \sin\alpha\; H^0_2, \\
H^0 &= \sin\alpha\; H^0_1 + \cos\alpha\; H^0_2. \\
\end{aligned}
\end{equation}
Denoting the ratio of the vacuum expectation values by $\langle H^0_2
\rangle/\langle H^0_1 \rangle = \tan\beta$, the Higgs-$W$-$W$ couplings read
\begin{equation}
\begin{aligned}
\frac{g(h^0WW)_{\rm THDM}}{g(HWW)_{\rm SM}} &= \cos(\beta-\alpha) \equiv\cos\xi,
\\
\frac{g(H^0WW)_{\rm THDM}}{g(HWW)_{\rm SM}} &= \sin(\beta-\alpha) \equiv\sin\xi,
\end{aligned}
\end{equation}
where $g(HWW)_{\rm SM}$ is the $HWW$ coupling in the SM. If
$\xi\equiv\beta-\alpha$ is non-negligible, both $h^0$ and $H^0$ play a role in
the unitarization of $WW$ scattering. As shown in Fig.~\ref{fig:2h}, the
interplay of the two Higgs bosons in this process can be tested experimentally
with the help of the MEM. In particular, a sample of just 100 events will be
sufficient to rule out the parameter region with large values of $\xi\gesim\pi/4$ and
$m_{H^0}\gesim 600\gev$ at about 90\% confidence level [assuming the data agrees with the SM]. 

\paragraph*{SILH:} The SILH paradigm encompasses a class of models with 
strong
dynamics at the scale $\Lambda \sim 4\pi f > 1\tev$ and a light composite Higgs
boson, which is a pseudo-Goldstone boson of some global symmetry \cite{silh}. As
a low-energy effective theory, it contains a light Higgs whose coupling to
$W$ and $Z$ are modified by a factor $1/\sqrt{1-c\, v^2/f^2}$, where $c$ is a ${\cal
O}(1)$ number. As a result, unitarization of high energy vector boson scattering
is not achieved by the light Higgs alone, but requires the presence of
additional heavy scalar and vector resonances, which emerge from the strong
sector. Here it is assumed that these resonance are beyond the reach of the LHC,
so that the only observable effect are the modified $hWW$ and $hZZ$
couplings\footnote{The explicit inclusion of the heavy resonances in the MEM does
not pose any conceptual problem, but due to the substantial amount of computing
time involved this is left for future work.},
which is equivalent to the limit $m_{H^0} \to \infty$ of the THDM.

\begin{figure}[t]
\centering%
\epsfig{figure=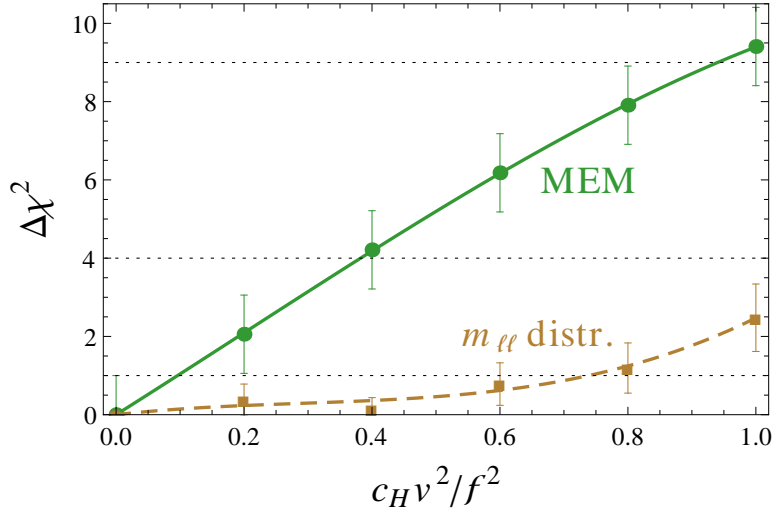, width=4in}%
\vspace{-1ex}
\caption{Statistical discrimination between different values of the deformation
parameter $c\, v^2/f^2$ in SILH models  from analyzing the process $pp \to
jjW^+W^+ \to jj\ell^+\ell^{\prime+}\nu_\ell\nu_{\ell'}$ using the MEM (circles)
vs.\ the di-lepton mass distribution (squares). The results are based
on 100 events simulated for $m_{H,\rm ref}=125\gev$, $c_H=0$, and $\sqrt{s}=14\tev$. The
error bars indicate the uncertainty from the event statistics.}
\label{fig:silh}
\end{figure}

Similar to the previous examples, the deformation parameter $c\, v^2/f^2$ can be constrained
by analyzing high energy $W^+W^+$ through the process $pp \to jjW^+W^+ \to
jj\ell^+\ell^{\prime+}\nu_\ell\nu_{\ell'}$. The output of the MEM as a function
of this parameter 
is shown in Fig.~\ref{fig:silh}, together with the results obtained from
analyzing the $m_{\ell\ell}$ distribution. It has been checked that the latter
are compatible with the numbers in Tab.~14 of Ref.~\cite{phantom11} within
statistical errors. As evident from the figure, the MEM leads to an improvement
of the sensitivity by more than one standard deviation.

\vspace{\bigskipamount}
\paragraph*{Conclusions:}
High energy vector boson scattering provides a unique window into the mechanism of
electroweak symmetry breaking, but it is difficult to analyze experimentally at
the LHC due to the small event yield. The Matrix Element Method (MEM), which is
an automated likelihood technique incorporating all relevant
event and theory information, significantly improves the sensitivity for this
process as compared with traditional analysis methods. This had been demonstrated
explicity here for high energy $W^+W^+$ scattering.
As concrete examples, the method has been applied to three characteristic
examples, the Standard Model with variable Higgs mass, the Two Higgs Doublet
Model, and the Strongly Interacting Light Higgs Model, but it can be adapted
straightforwardly to other models of electroweak symmetry breaking by
implementing the relevant matrix elements.

It is worth pointing out that the method does not rely on the observation of a
resonance. In fact, for high energy same-sign $WW$ scattering, the Higgs boson
or any other unitarity-restroring particle contributes only in the t-channel.
Consequently, the MEM will be useful even for a so-called ``nightmare'' scenario
with very broad resonances \cite{nightmare}.

The results presented here are based on a parton-level analysis. For a more
realistic picture, one needs to consider systematic uncertainties and detector
effects. The largest systematic error is related to the measurement of the jet
energy, which can be taken into account by incorporating jet smearing functions
and by treating the overall jet energy scale as a free parameter in the fit
\cite{memtev1}. While this leads to a substantial increase in computing time,
the senstivity of the MEM is not significantly reduced. 

A potentially important reducible background arises from events in
which a $W$ boson (which decays to an electron or positron together with
the corresponding neutrino) is produced with three associated jets, and one of the
jets is incorrectly identified as an electron or positron.  For
reasonable values of the rate at which jets are incorrectly
reconstructed as electrons ($\sim 10^{-4}$), a preliminary analysis
suggests that the rate for this process may be $\sim 15 \%$ of the
signal rate.  More exhaustive studies by the LHC detector
collaborations are necessary for a precise determination of
the significance of this background, though it can probably be reduced
by imposing more stringent criteria in electron reconstruction.

Other systematic effects
include next-to-leading order QCD corrections, which are small for vector boson
fusion processes \cite{nlo}, and uncertainties in the quark and antiquark parton
distribution functions (PDFs). While PDF errors for $WWjj$ production are
already at the level of a few percent \cite{pdferr}, the availability of
substantial LHC data should reduce these errors further, due to a better
understanding of the light quark flavor distribution in the relevant range of
$x$, and different experimental systematics compared to deep inelastic
scattering. Therefore it is expected that the effectiveness of the MEM will not
be significantly reduced by systematic errors.

\paragraph*{Acknowledgments:} The authors would like to thank J.~Alwall 
for
helpful communications.
AF gratefully acknowledges the warm hospitality at
the Michigan Center for Theoretical Physics during part of this
project.  JG likewise thanks the Aspen Center for Physics (funded by NSF Grant \#1066293)
for their hospitality and the SLAC
National Accelerator Laboratory for the use of computing resources.  
This work was partially supported by the National Science Foundation
under grant PHY-1212635 and by the Department of Energy under grants
DE-AC02-06CH11357, DE-FG02-91ER40684, and DE-FG02-97ER41029.


\renewcommand{\refname}



{\normalsize References}
\begin{thebibliography}{99}

\bibitem{higgs1}
  G.~Aad {\it et al.}  [ATLAS Collaboration],
  Phys.\ Lett.\ B {\bf 716}, 1 (2012);
  S.~Chatrchyan {\it et al.}  [CMS Collaboration],
  Phys.\ Lett.\ B {\bf 716}, 30 (2012).

\bibitem{history}
  M.~J.~Duncan, G.~L.~Kane and W.~W.~Repko,
  Nucl.\ Phys.\ B {\bf 272}, 517 (1986);
  D.~A.~Dicus and R.~Vega,
  Phys.\ Rev.\ Lett.\  {\bf 57}, 1110 (1986);
  R.~Kleiss and W.~J.~Stirling,
  Phys.\ Lett.\ B {\bf 200}, 193 (1988);
  V.~D.~Barger, K.~Cheung, T.~Han and R.~J.~N.~Phillips,
  Phys.\ Rev.\ D {\bf 42}, 3052 (1990);
  U.~Baur and E.~W.~N.~Glover, 
  Phys.\ Lett.\ B {\bf 252}, 683 (1990);
  D.~A.~Dicus, J.~F.~Gunion and R.~Vega,
  Phys.\ Lett.\ B {\bf 258}, 475 (1991);
  D.~A.~Dicus, J.~F.~Gunion, L.~H.~Orr and R.~Vega,
  Nucl.\ Phys.\ B {\bf 377}, 31 (1992);
  J.~Bagger {\it et al.}, 
  Phys.\ Rev.\ D {\bf 49}, 1246 (1994) and
  D {\bf 52}, 3878 (1995);
  K.~Iordanidis and D.~Zeppenfeld,
  Phys.\ Rev.\ D {\bf 57}, 3072 (1998);
  J.~M.~Butterworth, B.~E.~Cox and J.~R.~Forshaw,
  Phys.\ Rev.\ D {\bf 65}, 096014 (2002);
  A.~Alboteanu, W.~Kilian and J.~Reuter,
  JHEP {\bf 0811}, 010 (2008);
  C.~Englert, B.~J\"ager, M.~Worek and D.~Zeppenfeld,
  Phys.\ Rev.\ D {\bf 80}, 035027 (2009);
  A.~Ballestrero, G.~Bevilacqua and E.~Maina,
  JHEP {\bf 0905}, 015 (2009);
  A.~Ballestrero, G.~Bevilacqua, D.~B.~Franzosi and E.~Maina,
  JHEP {\bf 0911}, 126 (2009);
  G.~Aad {\it et al.}  [ATLAS Collaboration],
  arXiv:0901.0512 [hep-ex].

\bibitem{phantom11}
A.~Ballestrero, D.~B.~Franzosi and E.~Maina,
  JHEP {\bf 1106}, 013 (2011).

\bibitem{wwpol}
K.~Doroba {\it et al.},
  Phys.\ Rev.\ D {\bf 86}, 036011 (2012).

\bibitem{matrix}
  K.~Kondo,
  J.\ Phys.\ Soc.\ Jap.\  {\bf 57}, 4126 (1988) and
  {\bf 60}, 836 (1991);
  R.~H.~Dalitz and G.~R.~Goldstein,
  Phys.\ Rev.\  D {\bf 45}, 1531 (1992).

\bibitem{memtev1}
  B.~Abbott {\it et al.}  [D\O\ Collaboration],
  Phys.\ Rev.\  D {\bf 60}, 052001 (1999);
  V.~M.~Abazov {\it et al.}  [D\O\ Collaboration],
  Nature {\bf 429}, 638 (2004).

\bibitem{memtev2}
  A.~Abulencia {\it et al.}  [CDF Collaboration],
  Phys.\ Rev.\ D {\bf 75}, 031105 (2007);
  F.~Fiedler, A.~Grohsjean, P.~Haefner and P.~Schieferdecker,
  Nucl.\ Instrum.\ Meth.\  A {\bf 624}, 203 (2010).

\bibitem{singletop}
  V.~M.~Abazov {\it et al.}  [D\O\ Collaboration],
  Phys.\ Rev.\  D {\bf 78}, 012005 (2008);
  T.~Aaltonen {\it et al.}  [CDF Collaboration],
  Phys.\ Rev.\ Lett.\  {\bf 101}, 252001 (2008).

\bibitem{memhiggs}
S.-C.~Hsu {\it et al.}  [CDF Collaboration], CDF note 8774 (2007);
  T.~Aaltonen {\it et al.}  [CDF Collaboration],
  Phys.\ Rev.\ D {\bf 80}, 071101 (2009);
J.~Therhaag, Diplom thesis, University of Bonn, BONN-IB-2009-03 (2009);
  D.~E.~Soper and M.~Spannowsky,
  Phys.\ Rev.\ D {\bf 84}, 074002 (2011);
  J.~S.~Gainer, K.~Kumar, I.~Low and R.~Vega-Morales,
  JHEP {\bf 1111}, 027 (2011);
  J.~S.~Gainer, W.-Y.~Keung, I.~Low and P.~Schwaller,
  Phys.\ Rev.\ D {\bf 86}, 033010 (2012);
  P.~Avery {\it et al.},
  Phys.\ Rev.\ D {\bf 87}, 055006 (2013);
 J.~R.~Andersen, C.~Englert and M.~Spannowsky,
  Phys.\ Rev.\ D {\bf 87}, 015019 (2013).

\bibitem{memnlo}
  J.~M.~Campbell, W.~T.~Giele and C.~Williams,
  JHEP {\bf 1211}, 043 (2012).

\bibitem{lhcll}
  J.~Alwall, A.~Freitas and O.~Mattelaer,
  AIP Conf.\ Proc.\  {\bf 1200}, 442 (2010);
  C.-Y.~Chen and A.~Freitas,
  JHEP {\bf 1102}, 002 (2011).

\bibitem{madweight}
  P.~Artoisenet, V.~Lema\^itre, F.~Maltoni and O.~Mattelaer,
  JHEP {\bf 1012}, 068 (2010).

\bibitem{memisr}
  J.~Alwall, A.~Freitas and O.~Mattelaer,
  Phys.\ Rev.\ D {\bf 83}, 074010 (2011).

\bibitem{feynarts}
  T.~Hahn,
  Comput.\ Phys.\ Commun.\  {\bf 140}, 418 (2001).
  
\bibitem{madgraph}
  J.~Alwall, M.~Herquet, F.~Maltoni, O.~Mattelaer and T.~Stelzer,
  JHEP {\bf 1106}, 128 (2011).

\bibitem{thdm}
  W.~Altmannshofer, S.~Gori and G.~D.~Kribs,
  Phys.\ Rev.\ D {\bf 86}, 115009 (2012);
  S.~Chang, S.~K.~Kang, J.-P.~Lee, K.~Y.~Lee, S.~C.~Park and J.~Song,
  JHEP {\bf 1305}, 075 (2013);
  Y.~Bai, V.~Barger, L.~L.~Everett and G.~Shaughnessy,
  arXiv:1210.4922 [hep-ph].

\bibitem{silh}
  G.~F.~Giudice, C.~Grojean, A.~Pomarol and R.~Rattazzi,
  JHEP {\bf 0706}, 045 (2007).

\bibitem{nightmare}
  A.~Falkowski, C.~Grojean, A.~Kaminska, S.~Pokorski and A.~Weiler,
  JHEP {\bf 1111}, 028 (2011).

\bibitem{nlo}
  B.~J\"ager, C.~Oleari and D.~Zeppenfeld,
  Phys.\ Rev.\ D {\bf 80}, 034022 (2009).

\bibitem{pdferr}
  P.~Bolzoni, F.~Maltoni, S.-O.~Moch and M.~Zaro,
  Phys.\ Rev.\ D {\bf 85}, 035002 (2012).

\end{thebibliography}
\end{document}